\documentclass{article}
\usepackage{caption}

\usepackage[english]{babel}
\usepackage[letterpaper,top=2cm,bottom=2cm,left=2cm,right=2cm,marginparwidth=1.75cm]{geometry}
\usepackage{fancyhdr}

\usepackage{graphicx}
\usepackage[colorlinks=true, allcolors=blue]{hyperref}

\begin{document}
\title{Data quality control system and long-term performance monitor of the LHAASO-KM2A}
\author{\thanks{Corresponding author, B.W. Hou: houbw@ihep.ac.cn,  S. Wu: wusha@ihep.ac.cn, S.Z. Chen: chensz@ihep.ac.cn} }

\date{}
\maketitle
Zhen Cao$^{1,2,3}$,
F. Aharonian$^{4,5}$,
Axikegu$^{6}$,
Y.X. Bai$^{1,3}$,
Y.W. Bao$^{7}$,
D. Bastieri$^{8}$,
X.J. Bi$^{1,2,3}$,
Y.J. Bi$^{1,3}$,
W. Bian$^{9}$,
A.V. Bukevich$^{10}$
% ...
(The LHAASO Collaboration) \\
$^{1}$ Key Laboratory of Particle Astrophysics \& Experimental Physics Division \& Computing Center, Institute of High Energy Physics, Chinese Academy of Sciences, 100049 Beijing, China\\
$^{2}$ University of Chinese Academy of Sciences, 100049 Beijing, China\\
$^{3}$ TIANFU Cosmic Ray Research Center, Chengdu, Sichuan,  China\\
$^{4}$ Dublin Institute for Advanced Studies, 31 Fitzwilliam Place, 2 Dublin, Ireland \\
$^{5}$ Max-Planck-Institut for Nuclear Physics, P.O. Box 103980, 69029  Heidelberg, Germany\\
$^{6}$ School of Physical Science and Technology \&  School of Information Science and Technology, Southwest Jiaotong University, 610031 Chengdu, Sichuan, China\\
$^{7}$ School of Astronomy and Space Science, Nanjing University, 210023 Nanjing, Jiangsu, China\\
$^{8}$ Center for Astrophysics, Guangzhou University, 510006 Guangzhou, Guangdong, China\\
$^{9}$ Tsung-Dao Lee Institute \& School of Physics and Astronomy, Shanghai Jiao Tong University, 200240 Shanghai, China\\
$^{10}$ Institute for Nuclear Research of Russian Academy of Sciences, 117312 Moscow, Russia\\
% ...

\sloppy

\begin{abstract}
The KM2A is the largest sub-array of the Large High Altitude Air Shower Observatory (LHAASO). It consists of 5216 electromagnetic particle detectors (EDs) and 1188 muon detectors (MDs). The data recorded by the EDs and MDs are used to reconstruct primary information of cosmic ray and gamma-ray showers. This information is used for physical analysis in gamma-ray astronomy and cosmic ray physics. To ensure the reliability of the LHAASO-KM2A data, a three-level quality control system has been established. It is used to monitor the status of detector units, stability of reconstructed parameters and the performance of the array based on observations of the Crab Nebula and Moon shadow. This paper will introduce the control system and its application on the LHAASO-KM2A data collected from August 2021 to July 2023. During this period, the pointing and angular resolution of the array were stable. From the observations of the Moon shadow and Crab Nebula, the results achieved using the two methods are consistent with each other. According to the observation of the Crab Nebula at energies from 25 TeV to 100 TeV, the time averaged pointing errors are estimated to be  $-0.003^{\circ} \pm 0.005^{\circ}$ and $0.001^{\circ} \pm 0.006^{\circ}$ in the R.A. and Dec directions, respectively. 

\end{abstract}

\section{Introduction}
Cosmic ray, first discovered by Victor Hess in 1912 through balloon flight experiments \cite{Hess:1912srp}, are high-energy particles originating from deep space in the cosmos. They are primarily composed of various atomic nuclei, as well as a small amount of electrons and gamma-ray. The energy spectrum of cosmic rays mostly follows a power-law function from 10$^{9}$ to 10$^{20}$ eV. However, there is a notable change in the power-law index around several PeV (1PeV=10$^{15}$ eV), resulting in a "knee" structure in the spectrum. This structure contains important information about the origin of cosmic ray. It is generally believed that cosmic ray with energies around  and below the knee originate from astrophysical sources within our Galaxy. However, the exact nature of these sources and the mechanisms responsible for accelerating cosmic ray to such high energies remain a longstanding major scientific question. Ultra-high-energy (UHE) gamma-ray with energies greater than 0.1 PeV, unaffected by interstellar magnetic fields, are an important tool of searching for and identifying the PeV cosmic ray sources. A review about the progress of UHE gamma-ray astronomy can be found in \cite{chsz2024}. The LHAASO-KM2A, as the most sensitive UHE gamma-ray detector, will play a crucial role in unraveling the PeV cosmic ray sources.

The LHAASO is located on Haizi Mountain in Daocheng County, Sichuan Province, China, at an altitude of 4410 m above sea level. It is a composite extensive air shower (EAS) detector array consisting of three sub-arrays: the Kilometer Square Array (KM2A), the 78,000 m$^2$ Water Cherenkov Detector Array (WCDA), and the Wide Field-of-view atmospheric Cherenkov Telescope Array (WFCTA). The KM2A is primarily used for detecting gamma-ray with energies above 10 TeV. It was constructed and operated incrementally, with half of the array beginning scientific operations at the end of 2019, three-quarters of the array starting in December 2020, and full array operations commencing in July 20th, 2021. The operational duty cycle is close to 100\%. Based on the previous data from KM2A, the LHAASO collaboration has made several breakthrough advances in UHE gamma-ray astronomy. These include the discovery of the first dozen "PeVatrons", revealing a wonderful fact that the Milky Way is full of PeV particle accelerators \cite{2021Natur.594...33C}. Recently, the number of UHE gamma-ray sources has been increased to 43 \cite{2023firstcatalog}. Additionally, amazing progress was also achieved when measuring the brightest gamma-ray burst GRB 221009A \cite{2023GRB}, the Cygnus region \cite{2023Cygnus}, and the Galactic diffuse gamma-ray \cite{2023diffusegamma}. In the work about the observation of the Crab Nebula \cite{2021ChPhC..45b5002A}, with the data of half KM2A array, KM2A has demonstrated strong abilities, like the rejection power of cosmic ray induced showers is better than $4 \times 10^3$ at energies above 100 TeV, and the pointing error for gamma-rays is less than 0.1$^\circ$. These exciting physical achievements are all supported by high-quality experimental data, primarily achieved through the KM2A's three-level quality control system.

All detectors of the KM2A and their corresponding electronics are exposed to the harsh outdoor environment, which poses a challenge for the stable operation of the detectors, especially during the rainy season. Therefore, we need to monitor the status of over 6000 detectors, perform maintenance on those that are malfunctioning, and exclude the abnormal data from subsequent event reconstruction and physical analysis. In response to the data collected by the KM2A detectors, we have established an automated data transfer, calibration and reconstruction system. To ensure timely data reconstruction, each file is submitted to a separate CPU for reconstruction. We also need to monitor the quality of the reconstructed data and select high-quality data files for the entire collaboration in subsequent physics analysis. The stability of array performance including pointing accuracy, angular resolution, and detection efficiency is crucial. Therefore, we should also monitor the standard source to ensure the physical reliability of the entire dataset. This constitutes the third level quality control system of KM2A data. In this article, we will introduce the specific implementation of the three-level quality control system for KM2A data from August 2021 to July 2023, and present the data status during this period.

\section{The detector and data processing flow of KM2A}

\subsection{The KM2A detector}
KM2A is the largest sub-array of LHAASO. It contains 5216 electromagnetic particle detectors (EDs) and 1188 underground muon detectors (MDs), which are distributed in an area of 1.3 $\mathrm{km^2}$. Within a radius of 575 m from the center of the array, EDs are arranged with a spacing of 15 m, while MDs are distributed with a spacing of 30 m. In the outer ring region, the spacing of EDs is enlarged to 30 m, with no MDs.

The ED is a plastic scintillation detector. Each ED consists of 4 plastic scintillation tiles (0.25 $\mathrm{m^2}$ each) covered by a 0.5 cm thick lead plate to convert the gamma-ray into electron-positron pairs. When a high energy charged particle traverses the scintillator, it loses energy and excites the scintillation medium to emit scintillation photons. Each ED also includes a 1.5-inch photomultiplier tube (PMT) to convert the scintillation photons to electrical signals to be recorded. The detection efficiency of a typical ED is about 98\%. The time resolution is about 2 ns. The average single rate of an ED is about 1.7 kHz. The electrical signals recorded by EDs are used to reconstruct the primary information of the cosmic ray air shower, such as the primary direction, core location, and energy.

The MD is a super-pure water Cherenkov detector enclosed within a cylindrical concrete tank with an inner diameter of 6.8 m and a height of 1.2 m. An 8-inch PMT is installed at the center of the top of the tank to collect the Cherenkov light produced by high energy particles as they pass through the water. The detectors are covered by a layer of soil 2.5 meters thick, which are used to absorb the secondary electrons/positrons and gamma-rays in showers while let the muons pass through. The detection efficiency of a typical MD to muons is about 95\%. The time resolution of an MD is about 10 ns. The average single rate of an MD is about 8 kHz. The number of muons recorded by MDs are used to discriminate between gamma-ray and hadron induced showers.

More details about the ED and MD design can be found elsewhere \cite{He2018,2022ChPhC..46c0001M}. KM2A operates around the clock, since both EDs  and MDs can work during both day and night. The KM2A detectors  were constructed and merged into the data acquisition system (DAQ) in stages since February 2018. The KM2A with the full configuration has operated since July 20th, 2021. The trigger logic of KM2A for a shower requires at least 20 EDs firing within a window of 400 ns. The starting time of the window is called the trigger time. For each event, the DAQ records 5 $\mu$s of data before the trigger time and 5 $\mu$s of data after the trigger time, including all EDs and MDs that have signals. The event trigger rate of KM2A is about 2.5 kHz, and the typical data size recorded in one day is about 2.5 TB.

\subsection{The data processing flow}
\begin{figure*}
\centering
\includegraphics[width=16cm,height=6cm]{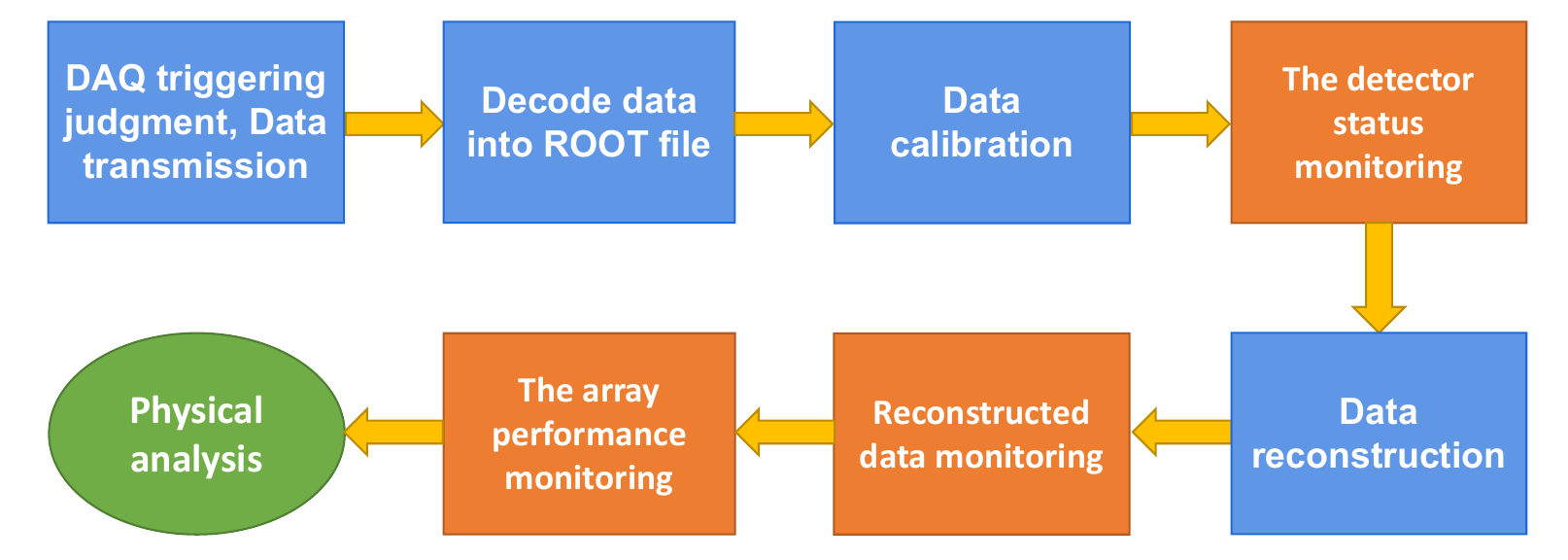}
\caption{The processing flow of the KM2A data. The process nodes marked with orange color indicate the three-level quality control system established for KM2A data.}
\label{fig1}
\end{figure*}

The data recorded by KM2A are stored in a binary file format, with the size of each file being approximately 1 GB. The daily data volume from KM2A is 2.5 TB, with approximately 2000-2500 files. These data files will be timely transmitted from the LHAASO site to the National HEP data center in Beijing. These data are used for physical analysis after the data processing, as shown in Fig.~\ref{fig1}. First, the binary data will be decoded into a ROOT file. Then, the arrival time and signal integrated charge of secondary particles recorded by EDs and MDs are calibrated using an offline method. Details about the calibration method can be found in \cite{PhysRevD.106.122004,LV201822}. After that, the states of all the detectors are monitored, picking out the abnormal detectors that will not be considered in the following reconstruction process. The arrival time and number of the particles recorded by normal EDs are used to reconstruct the primary information of the shower event, including the zenith angle ($\theta$) and azimuth angle ($\phi$), core position, number of electromagnetic particles (N$_e$), lateral distribution function and so on. The arrival time and number of the particles recorded by normal MDs are used to reconstruct the number of muons (N$_{\mu}$). The events with this information will be used for physical analysis both in gamma-ray astronomy and cosmic ray physics.

To ensure the quality of the data used for the physical analysis, the files with reconstructed data should also be monitored and the files with abnormal data should be filtered out. The final surviving normal data files will be used to monitor the performance of the KM2A through observations of the Crab Nebula and the Moon shadow. After this processing flow, the surviving normal files will be opened to the LHAASO collaboration for formal physical analysis. The pipeline of the data processing is shown in Fig.~\ref{fig1}, where the process nodes marked with orange color indicate the three-level quality control system established for LHAASO-KM2A data. All the physical results published by the LHAASO collaboration that relate to KM2A data are achieved using the data produced by this control system. In the following section, we will introduce the details of this control system.

\section{ The three-level quality control system of KM2A data}
The data recorded by EDs and MDs are used to reconstruct primary information of showers. Therefore, the first level is to monitor the detector status and filter out abnormal detectors, which will ensure only data recorded by normal detectors are used for event reconstruction. The second level is to monitor reconstructed data and filter out any abnormal data, which will ensure only reconstructed data collected during normal periods are used for physical analysis. The third level is to monitor the performance of the KM2A through observations of the Crab Nebula and the Moon shadow, which will ensure the stability of the KM2A performance. To introduce the details of this three-level quality control system, the KM2A data collected from August 2021 to July 2023 are analyzed in this section.

\subsection{The detector status monitoring}
All detectors of the KM2A and their corresponding electronics are exposed to the harsh outdoor environment. The maximum temperature difference in Daocheng can be more than 50 $^{\circ}$C, with a maximum exceeding 20$^{\circ}$C and minimum dropping below -35$^{\circ}$C. The rainy season from June to October brings precipitation levels of up to 515.2 mm. All these pose a challenge for the stable operation of the detectors, especially during the high lightning season \cite{Yan2023}. Extreme low temperatures can cause abnormal noise in detector electronics, and some of the detectors are struck by lightning every rainy season. Therefore, they have a certain probability to go wrong. This will result in abnormal data recorded by some detectors and the direct impact is on the count rate of the detectors. It is worth noting that, we have previously established a similar monitoring system mainly for the maintenance of the detectors \cite{Wang19}. That system primarily relied on hourly data, however, the monitoring presented here is based on each data file, corresponding to 40 seconds of data.

The count rate refers to the frequency at which a detector is triggered, which can be used to reflect the noise level of the detector and the environmental background level under normal conditions. A high count rate is generally due to significant electronic noise, while a low count rate is generally due to insufficient detection efficiency or loss of data for some time periods. Most of the signals from all EDs and MDs during the period from 5 $\mu$s to 1 $\mu$s before the trigger time are due to random particles and noise signals. Therefore, the data during this period are used to estimate the count rate of each detector. 

\begin{figure*}
\centering
\includegraphics[width=8cm,height=6cm]{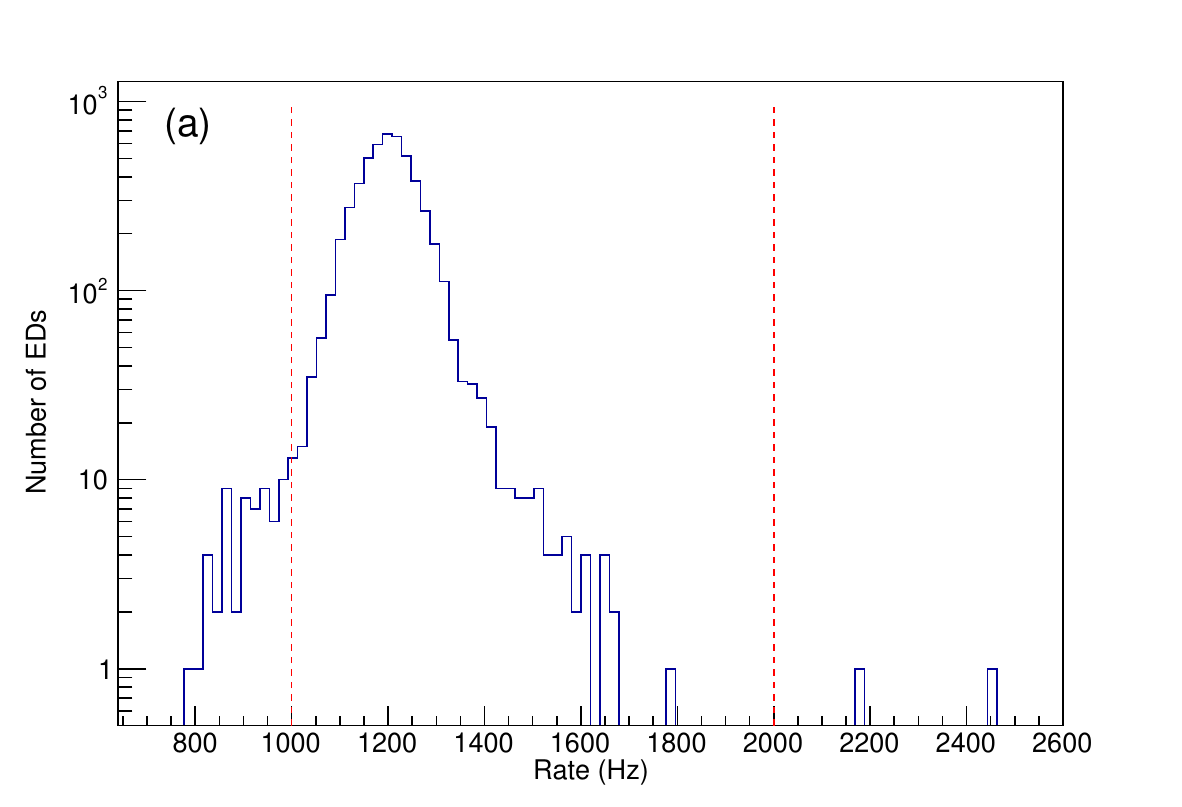}
\includegraphics[width=8cm,height=6cm]{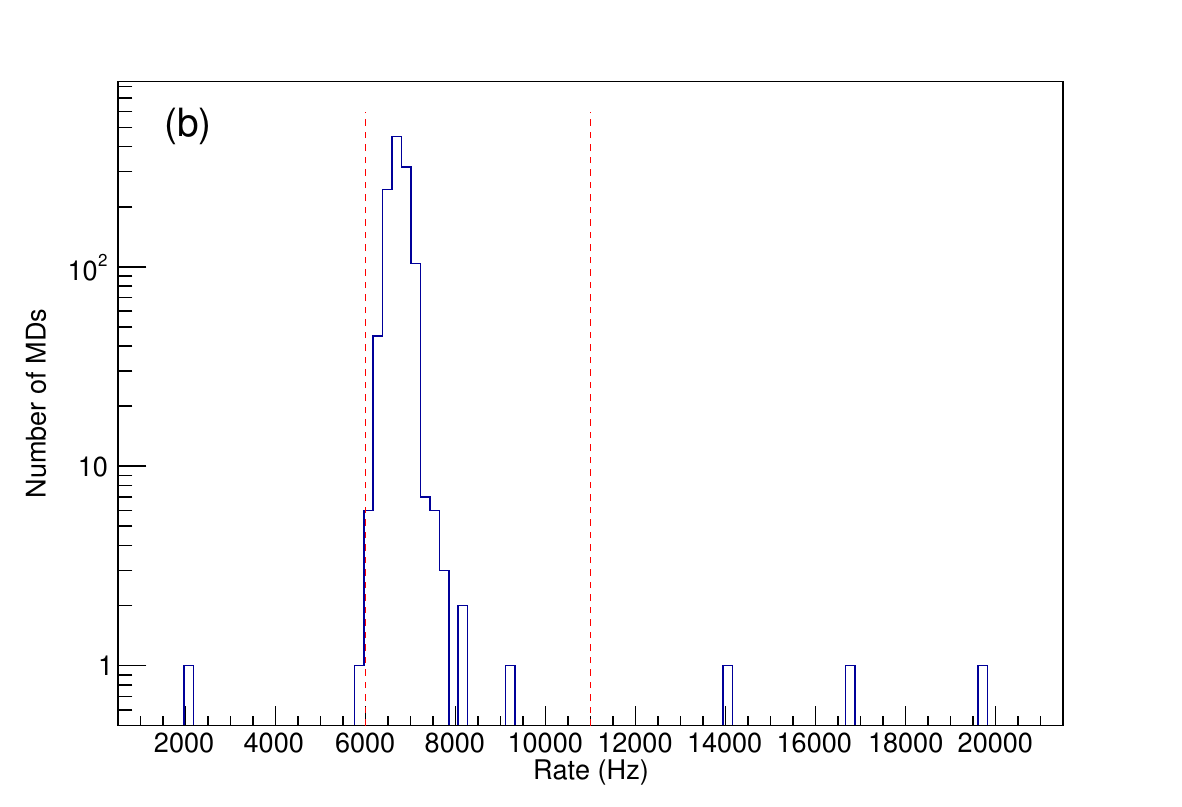}
\caption{(a): The distribution of EDs count rate. (b): The distribution of MDs count rate. A data file collected on 7th Feb 2022 is used for this figure.}
\label{fig2}
\end{figure*}

Basing on each data file, we can obtained the count rate of each ED and MD as shown in Fig \ref{fig2} for example. According to the detector efficiency and statistical error, the typical count rate of a ED should be 1-2 kHz, while a typical MD rate ranges from 6-11 kHz. Detectors beyond this range are labeled as abnormal detectors. The hits recorded by these abnormal detectors will not be used in the following reconstruction process. In this example file on 7th Feb 2022, 66 out of 5216 EDs and 5 out of 1188  MDs were flagged as abnormal detectors.
After the above screening work, we can eliminate the impact of extreme environments on the detectors.

\subsection{Reconstructed data monitoring and selection}
Following the detector status monitoring, the data recorded by the abnormal detectors are isolated during the event reconstruction phase. Nonetheless, the resultant reconstructed data might not meet the necessary quality for subsequent physical analysis. The second level would focuses on monitoring reconstructed data and filtering out any abnormal data. If there is a large fraction of abnormal detectors, the performance of the array diverges from the normal values. Although we can conduct physical analysis based on partial detector array data from KM2A, this will requires us to employ more complex data analysis methods. In order to maintain stability in detector performance and simplify the data analysis process, we generally require that the percentage of normal EDs and MDs (denoted as N$_{\rm ED}$ and N$_{\rm MD}$, respectively) reaches 95\% or more. Otherwise, we will label such data as abnormal, and they will not be used for physical analysis. Of course, if there are short-term important burst phenomena during the period of the excluded data, we will still analyze them using more complex analysis methods. In addition to requirement of at least 95\% good detectors, we will also monitor the average values of the following parameters for each data file:
\begin{itemize}
  \item N$_e$ and N$_{\mu}$: N$_e$ is the number of electromagnetic particles recorded by EDs. N$_{\mu}$ is the number of muons recorded by MDs. N$_e$ will be used in the energy reconstruction, and the ratio of N$_e$ and N$_{\mu}$ will be used for particle identification. These parameters also serve as indicators of the detectors for particle count. 
  
  \item $\theta$ and $\phi$:
$\theta$ is the zenith angle and $\phi$ is the azimuth angle. $\theta$ and $\phi$ represent the source direction of the event, and are important parameters in gamma-ray astronomy. These parameters also serve as indicators of the event direction reconstruction.

  \item $\chi^{2}$:
$\chi^{2}$ is the time residual squared during shower front fitting for direction reconstruction which reflects the stability of detector time measurement. This parameter also serves as an indicator of detector time resolution and time calibration. 

 \item $R$:
This parameter is the ratio of events with few muons. This parameter serves as an indicator of the discrimination power between cosmic ray and gamma-ray. In our subsequent physical analysis, specific conditions may be employed to help select events generated by gamma-ray. During this phase, we employ a relatively lenient screening criterion aiming to preserve a larger portion of the cosmic ray events.
\end{itemize}

The events of KM2A being triggered are basically isotropic with only a 0.1$\%$ level of anisotropy, and do not vary with time. Therefore, the distribution of the above parameters in each file should be almost the same and remain stable over time, making them suitable for data quality monitoring. However, in actual measurements, there are slight modulation effects on these variables on a daily and yearly basis due to meteorological effects and atmospheric changes. Additionally, the variation in the number of normal detectors also affects the corresponding parameter changes. Therefore, when using the above parameters to monitor reconstructed data and select abnormal data, we primarily identify data files that deviate significantly from the distribution based on the daily behavior of these parameters over time.

Our initial step involves applying a broad condition to screen out files exhibiting extreme anomalies. Subsequently, Gaussian fitting is conducted on the remaining files to derive the mean and standard deviation. Considering the daily volume of generated data files (2000-2500), we have instituted a 4$\sigma$ filtering criterion. This stringent condition aims to exclude aberrant data while mitigating the potential risk of erroneously discarding more than one normal file per day. Upon completion of the filtering process, a compilation of normal data files is generated, deemed suitable for further analysis in the physical domain. Upon completion of detector status monitoring and reconstruction data quality monitoring, a total of 1,700,628 reconstructed data files spanning from July 20, 2021, to July 31, 2023, corresponding to the initial 2 years of full array operation, were obtained. Applying all the filtering criteria led to the exclusion of 30,066 files, accounting for approximately 1.77\% of the total dataset.

\begin{figure*}
\centering
\includegraphics[width=16cm,height=6cm]{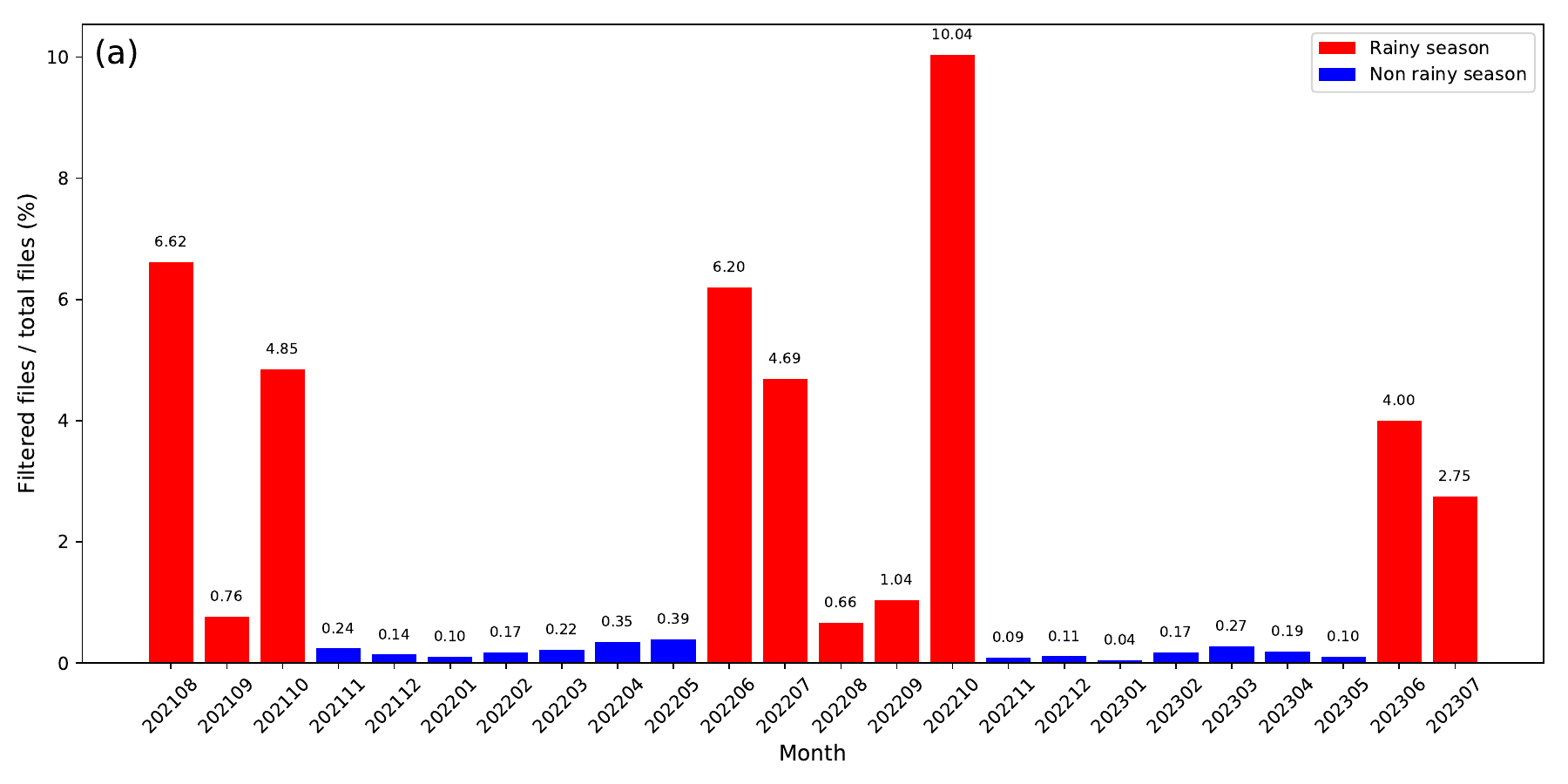}
\includegraphics[width=10cm,height=6cm]{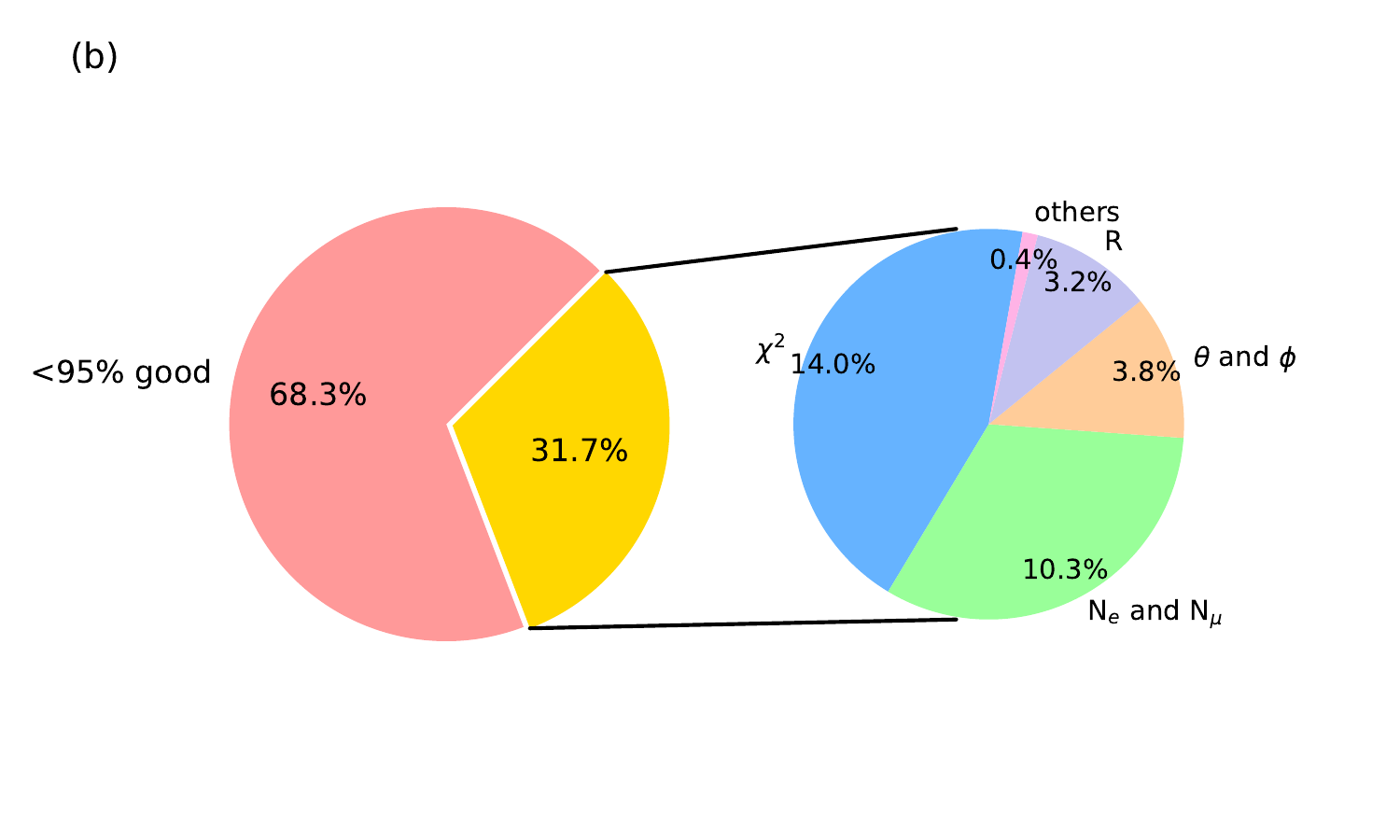}
\includegraphics[width=6cm,height=6cm]{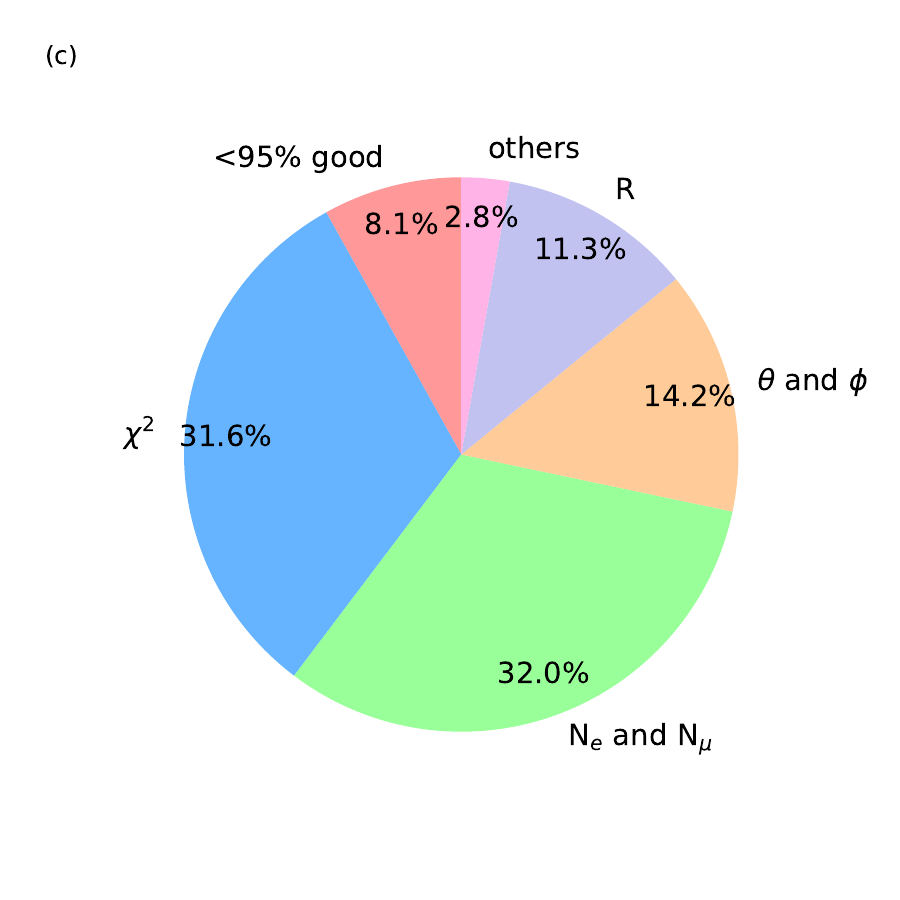}
\caption{(a): The percentage of data files excluded based on different criteria. (b): The ratio of excluded data files reference to the primary total data files in each month during the period from August 2021 to July 2023. (c): The ratio of excluded data files in non-rain season.}
\label{fig4}
\end{figure*}

Fig~\ref{fig4} (a) shows the variation of the exclusion rate in the monthly data. It is evident that there is a seasonal effect, with the exclusion rate being significantly higher from June to October each year, which is the rainy season. During the annual rainy season, the electronics of the ED and MD detectors are susceptible to lightning strikes, occurring approximately once or twice a year. Each lightning strike can result in damage to tens to hundreds of detector electronics or power supplies. Due to the extended duration of repairs, the normal detectors are generally unable to meet the requirement of 95\% for good EDs and MDs during this period, resulting in a relatively high data exclusion rate, reaching as high as 10\% in October 2022. Additionally, because rainwater contains radioactive radon elements, this can lead to an increase in the background noise rate of the ED detectors, thereby affecting the triggering efficiency of the array and causing changes in various parameters, resulting in an increased data exclusion rate. When there are no lightning strikes, the data exclusion rate is around 1\% during the rainy season. During the non-rainy season, the data exclusion rate ranges from 0.04\% to 0.4\% for a total of 1855 files, mainly due to abnormal N$_e$ and N$_{\mu}$ or $\chi^{2}$.

Among the excluded files, 68.3\% of the files were excluded because they had < 95\% good EDs and MDs, 14.0\% were due to $\chi^{2}$, 10.3\% were due to N$_e$ and N$_{\mu}$, 3.8\% were due to $\theta$ and $\phi$, 3.2\% were due to R, and 0.4\% were due to other reasons. The distribution is also shown in Fig~\ref{fig4} (b). During the non-rainy season shown in Fig~\ref{fig4} (c), 32.0\% of the files were excluded due to N$_e$ and N$_{\mu}$, 31.6\% were due to $\chi^{2}$, 14.2\% were due to $\theta$ and $\phi$, 11.3\% were due to R, 8.1\% were due to <95\% good EDs and MDs, and 2.8\% were due to other reasons.

\begin{figure*}
\centering
\includegraphics[width=16cm,height=12cm]{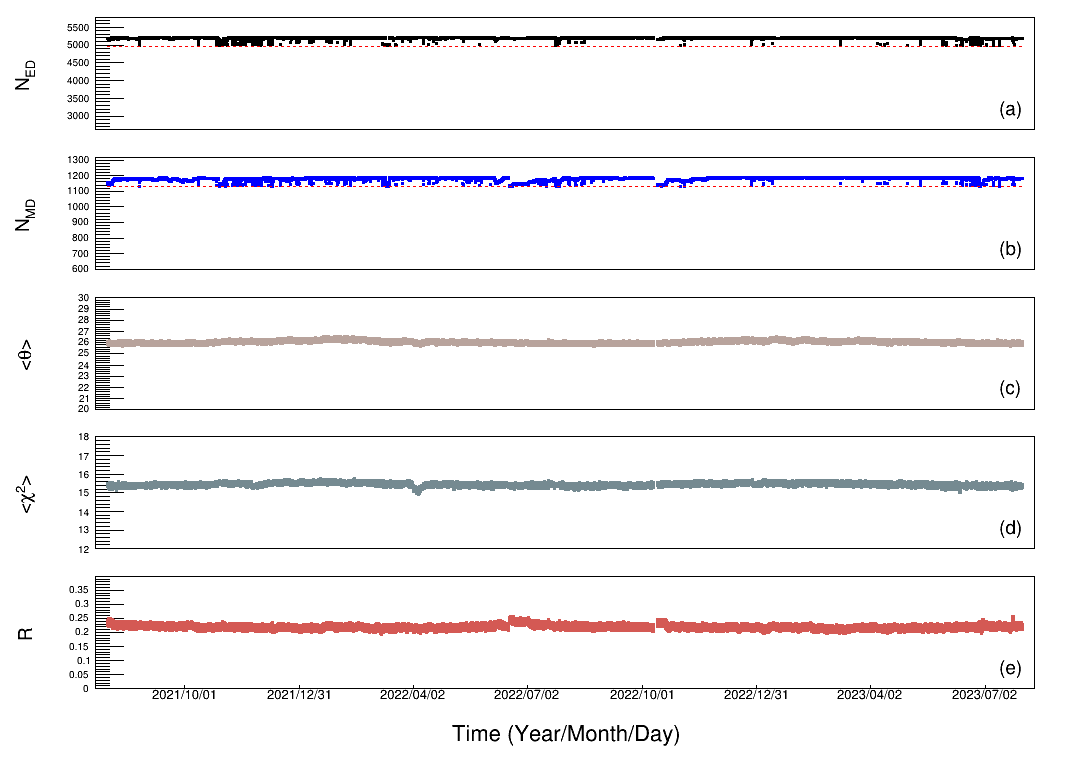}
\caption{The parameters N$_{ED}$ and N$_{MD}$ and three average values for each data file during the period from July 20th, 2021 to 31st July 2023. The dotted lines indicate the requirement of 95\% good detectors.}
\label{fig5}
\end{figure*}

After monitoring and filtering out any abnormal data from the reconstructed data, the remaining qualified data will be used for subsequent physical analysis, including cosmic ray physics, gamma-ray astronomy, and so on. To provide an overview of the long-term data status, Fig~\ref{fig5} shows the variation of the parameters N$_{\rm ED}$ and N$_{\rm MD}$ and three average values for each data file. This R increases when the number of normal MDs decreases. This is primarily due to the fact that the gamma-ray/cosmic ray discrimination power is mainly determined by the detection efficiency of the muon detectors. For the majority of the time, nearly 99\% of the detectors are in a normal state and this R remains stable.

\subsection{The array performance monitoring}
Before using the reconstructed data for physical analysis, the final step to ensure data quality is to monitor the performance of the KM2A by observing a gamma-ray source, the Crab Nebula and a cosmic ray deficit source, the Moon shadow. In this work, we use two years of data collected from August 2021 to July 2023 by the KM2A full-array to analyze the Crab Nebula and Moon shadow. For background estimation, we employ the equal zenith angle method. The likelihood method presented in \cite{2017ICRC} is used to estimate the significance map, position, point spread function and so on.

\subsubsection{Observations of Crab Nebula }
The Crab Nebula, located approximately 2 kiloparsecs from Earth, is recognized as the brightest pulsar wind nebula in the northern celestial region. It is the remnant of a supernova explosion, which was recorded in Chinese and Japanese chronicles dating back to 1054 A.D.\cite{green2003}. The Crab Nebula is one of the most studied objects in the sky, and its emission spectrum extends from radio waves up to PeV gamma-ray \cite{2021Sci...373..425L}. Below a few GeV, the Crab Nebula has been observed to exhibit flaring behavior, but in the very high energy (VHE) band, it remains a constant source and is currently utilized as the standard reference by different instruments. The performance of the KM2A half-array on gamma-ray has been tested through the analysis of the Crab Nebula \cite{2021ChPhC..45b5002A}. 

\begin{figure*}
\centering
\includegraphics[width=7cm,height=6.5cm]{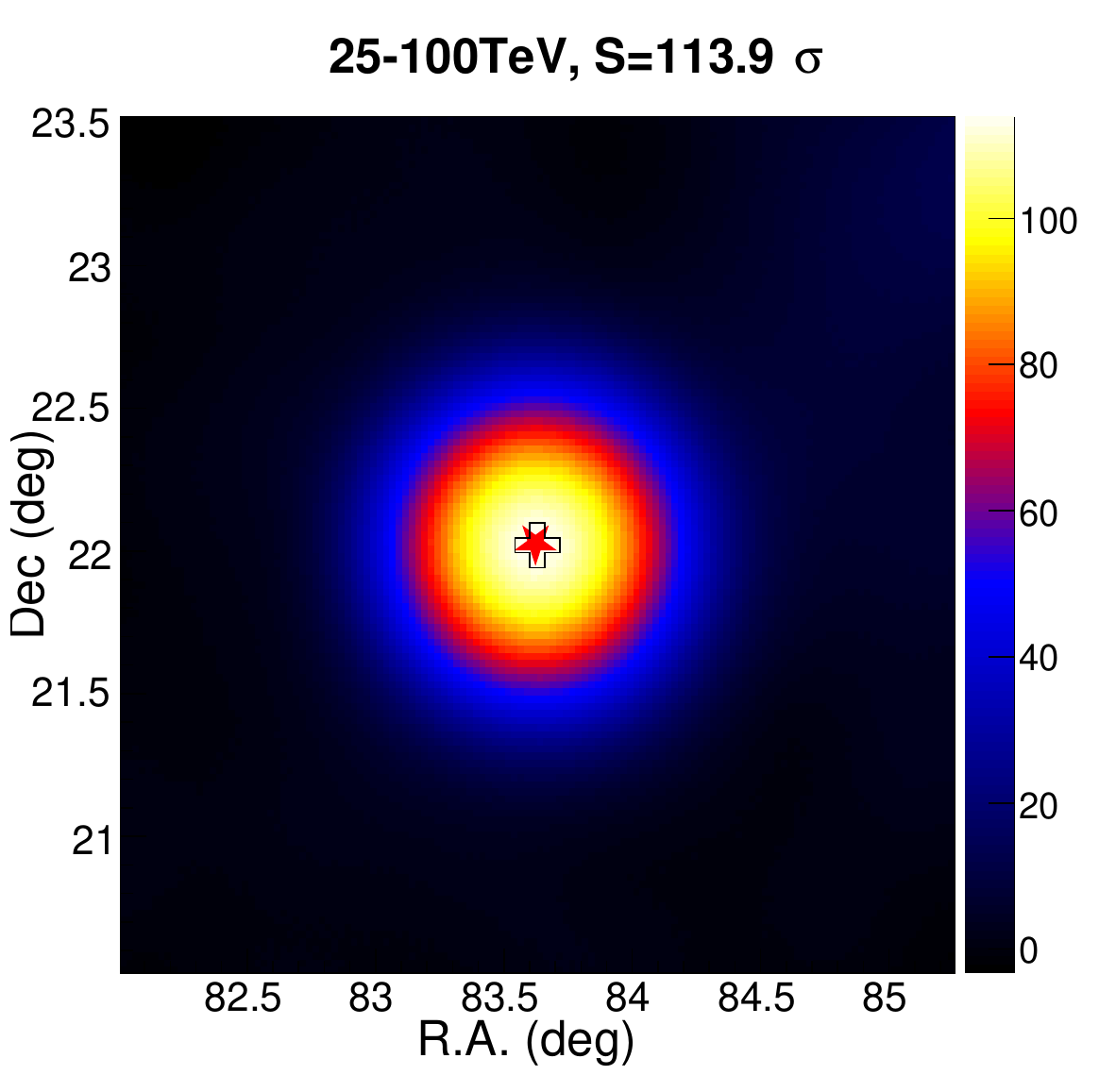}
\includegraphics[width=7cm,height=6.5cm]{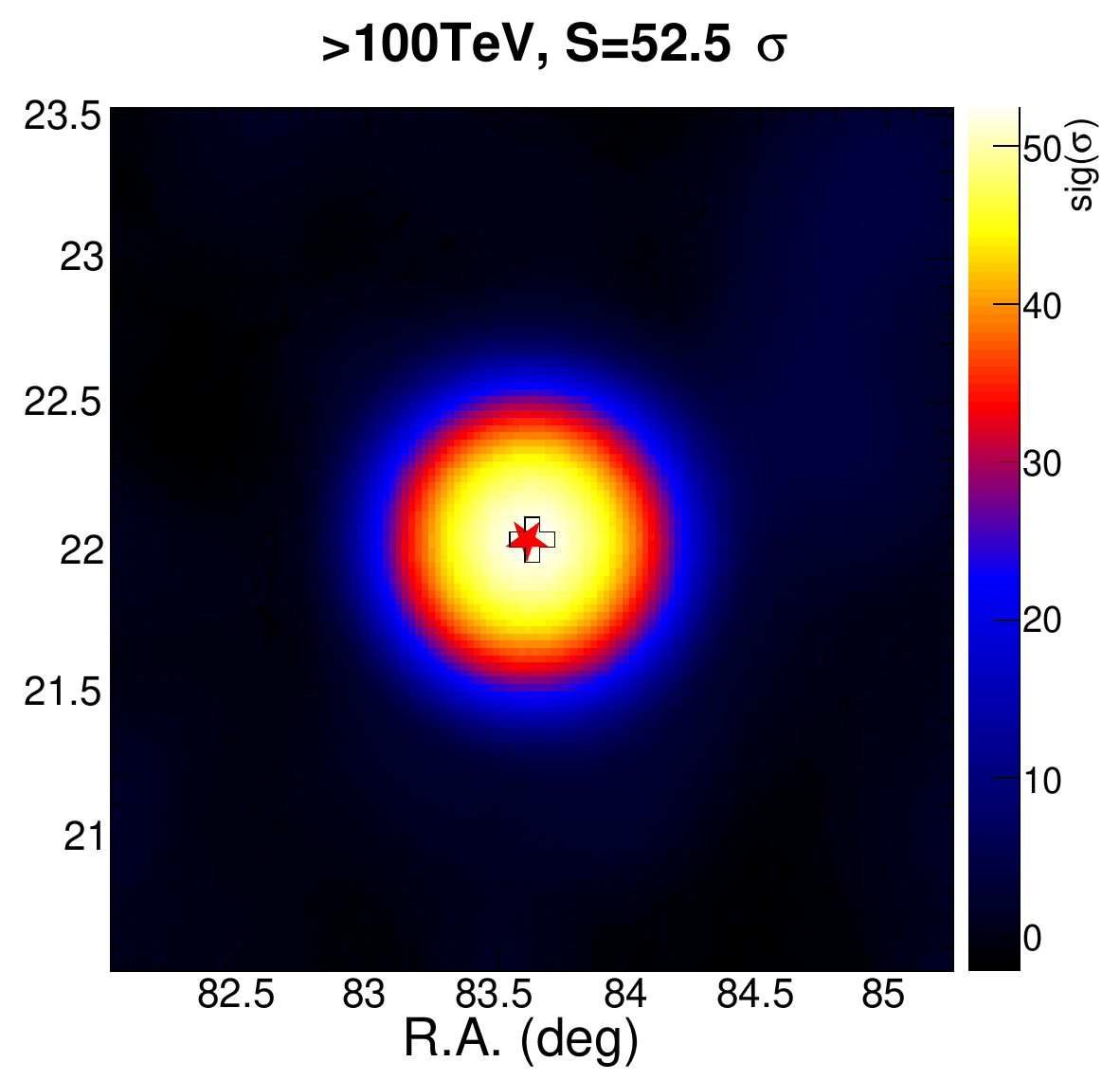}
\caption{Left: The significance map around the Crab Nebula at gamma-ray energies of 25-100 TeV. Right: The significance map around the Crab Nebula at gamma-ray energies above 100 TeV. The red star is the fitting position and the black cross is the expected position at R.A.=$83.63^\circ$, Dec=$22.02^\circ$.}
\label{fig6}
\end{figure*}

\begin{figure*}
\centering
\includegraphics[width=14cm,height=14cm]{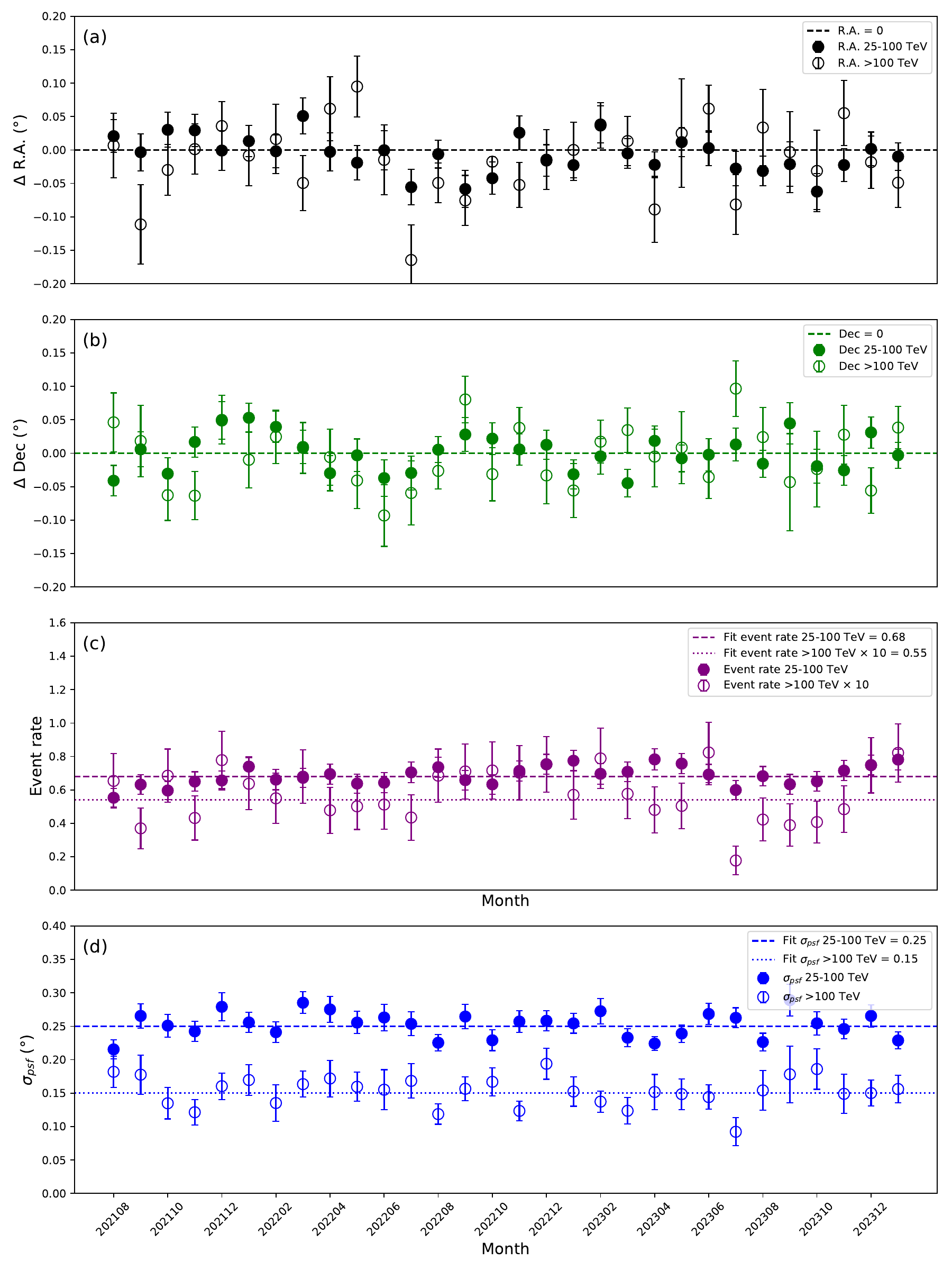}
\caption{The monthly measured centroid of the Crab Nebula reference to the expected position in R.A. (Panel a) and Dec (Panel b) directions as a function of time. The solid lines show constant values that fit the centroid for all times. Panel c: The monthly average signal event rate measured from the Crab Nebula direction as a function of time. Panel d: The monthly angular resolution obtained using the observation of Crab Nebula as function of the time.}
\label{fig7}
\end{figure*}

For this work, the significance map around the Crab Nebula is shown in Fig \ref{fig6}. The significance of the Crab is 113.9$\sigma$ in the energy band of 25-100 TeV and 52.5$\sigma$ above 100 TeV. Therefore, it is significant enough for us to conduct monthly monitoring of the performance of KM2A for gamma-ray detection. The monthly results are presented in Fig \ref{fig7}. According to this figure, the pointing errors in both R.A. and Dec direction are much less than $0.1^\circ$ at 25-100 TeV and less than $0.2^\circ$ at $>$100 TeV in all the months. Constant values are adopted to fit the points at 25-100 TeV, yielding that $\Delta R.A.$= $-0.003^{\circ} \pm 0.005^{\circ}$ and $\Delta Dec$= $0.001^{\circ} \pm 0.006^{\circ}$. The $\chi^2/ndf$ of the fitting are 26.66/23 and 32.62/23, respectively. The corresponding values at $>$100 TeV are $\Delta R.A.$= $-0.016^{\circ} \pm 0.007^{\circ}$ and $\Delta Dec$= $-0.002^{\circ} \pm 0.010^{\circ}$. The $\chi^2/ndf$ of the fitting are 40.39/23 and 35.13/23, respectively. So $0.008^\circ$ and $0.023^\circ$ can be seen as conservative upper limits for combined systematic and statistical errors at 25-100 TeV and $>$100 TeV.


To monitor the detection efficiency of KM2A, we estimate the event rate from the Crab Nebula, which is also shown in Fig \ref{fig7}. A constant value is adopted to fit the points at 25-100 TeV, yielding that $Rate$= $0.68 \pm 0.01$ per hour and the corresponding $\chi^2/ndf$ is 23.46/23. For the events at $>$100 TeV, the corresponding value is $Rate$= $0.054 \pm 0.004$ per hour and $\chi^2/ndf$ is 36.18/23. Therefore, the detection efficiency is also stable month by month.

Compared with the point spread function (PSF) of the KM2A detector, the intrinsic extension of the Crab Nebula is negligible. Therefore, the angular distribution of gamma-ray from the Crab Nebula can be directly used to estimate the detector angular resolution. The monthly angular resolution results are also shown in Fig \ref{fig7}. A constant value is adopted to fit the points at 25-100 TeV, yielding $\sigma_{PSF} $= $0.248^{\circ} \pm 0.004^{\circ}$ and the corresponding $\chi^2/ndf$ is 34.85/23. For the events at $>$100 TeV, the corresponding value is $\sigma_{PSF} $= $0.146^{\circ} \pm 0.005^{\circ}$ and $\chi^2/ndf$ is 29.67/23. Therefore, the angular resolution is stable month by month. So $0.252^\circ$ and $0.151^\circ$ are the upper limits for combined systematic and statistical errors at 25-100 TeV and $>$100 TeV.

\subsubsection{Observations of Moon shadow }

Cosmic ray arrive at Earth in an almost isotropic manner. However, as the Moon blocks any cosmic ray arriving from that direction in the sky, a deficit in the flux can be observed. Many EAS experiments have observed the Moon shadow, providing unique information on the performance of cosmic ray, including pointing and angular resolution \cite{YBJ2011}. Additionally, the shadow is displaced along the east-west direction from the Moon's actual position due to geomagnetic deflection, which can be used to determine the absolute energy scale of the primary cosmic ray. However, that part of the research goes beyond the scope of this article.

\begin{figure*}
\centering
\includegraphics[width=7cm,height=6.5cm]{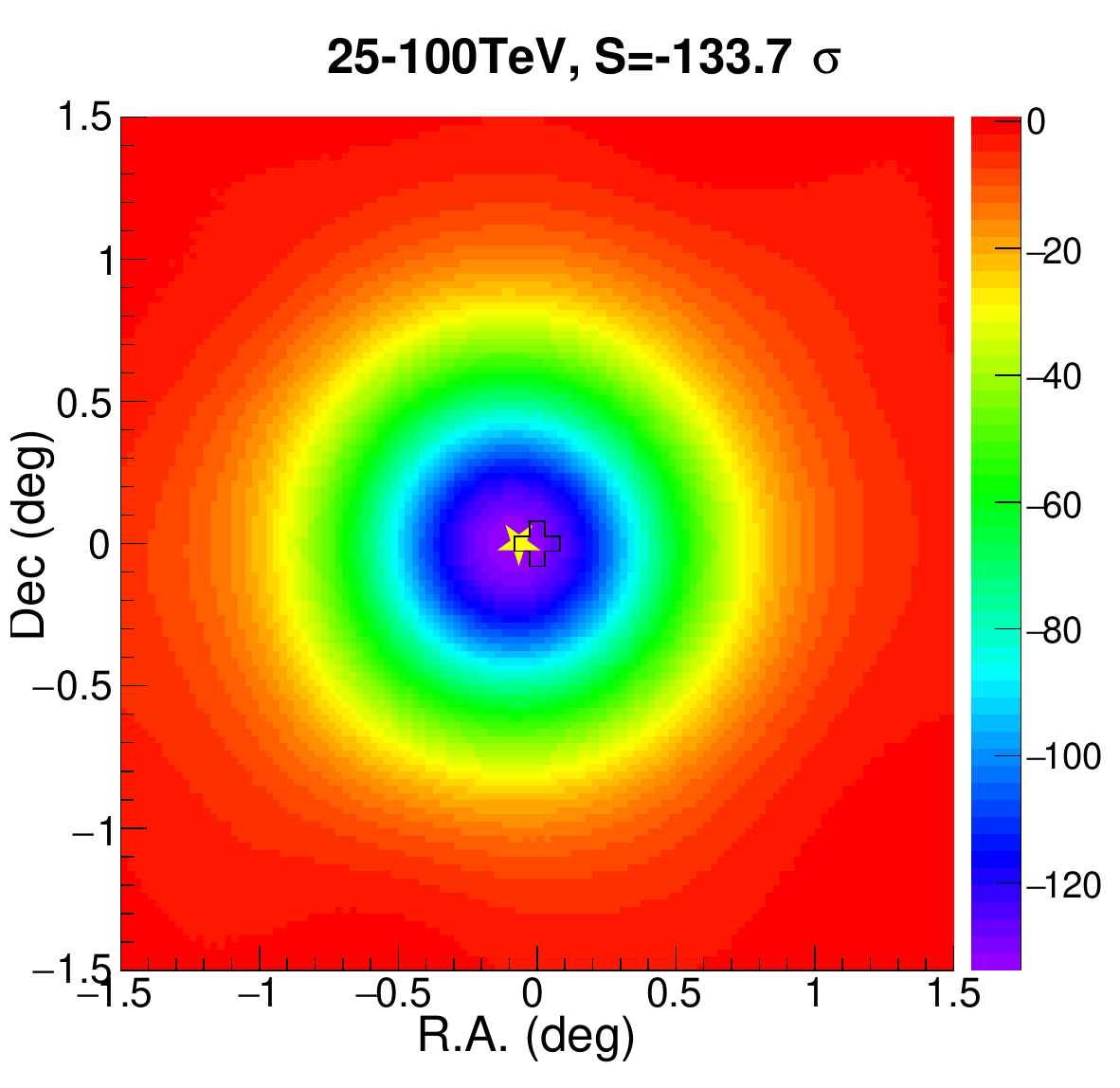}
\includegraphics[width=7cm,height=6.5cm]{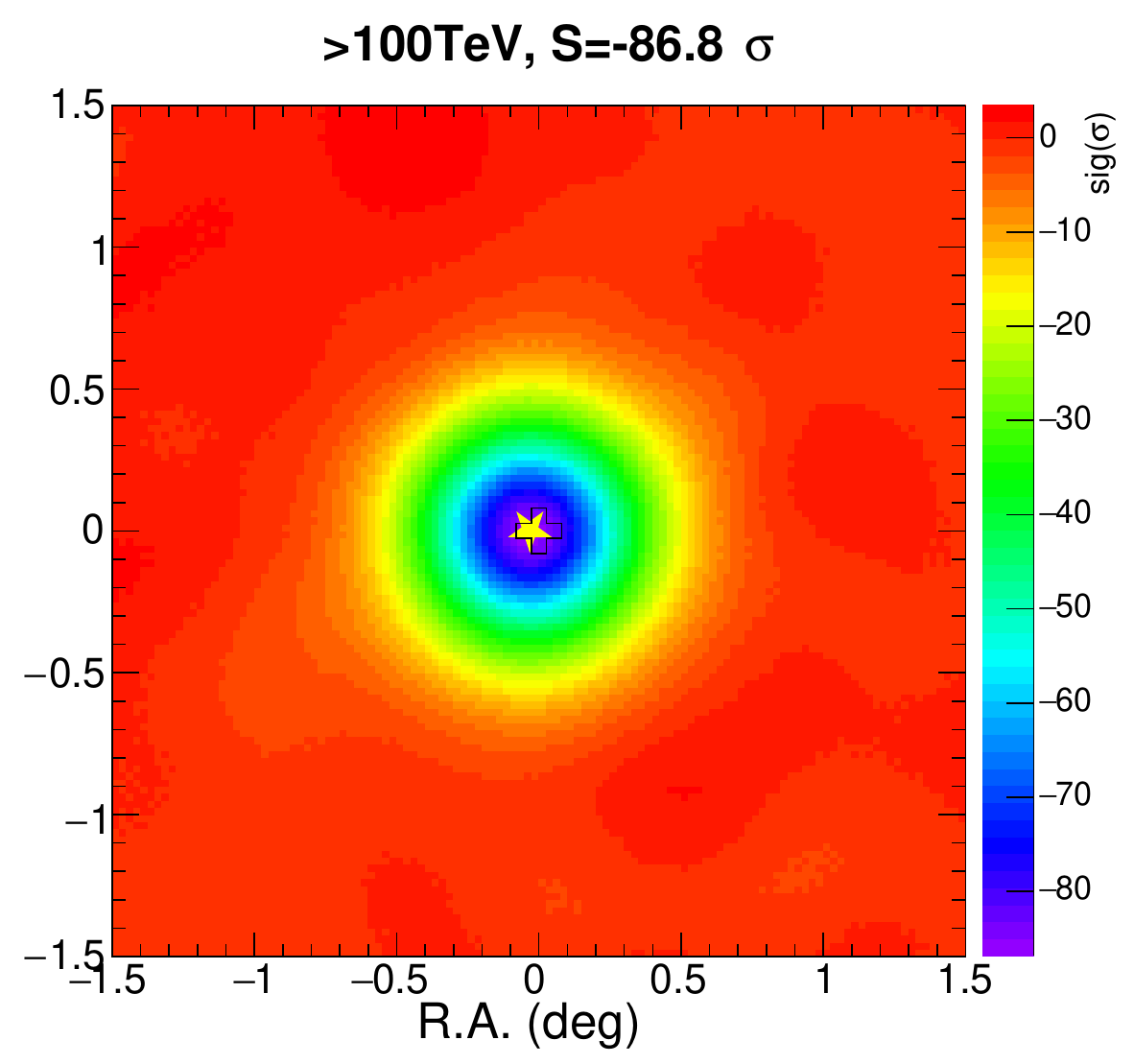}
\caption{Left: The significance map around the Moon at gamma-ray energies of 25-100 TeV. Right: The significance map around the Moon at gamma-ray energies above 100 TeV. The yellow star is the fitting position of the Moon shadow and the black cross is the expected Moon position.}
\label{fig8}
\end{figure*}

For this work, the significance map around the Moon shadow is shown in Fig \ref{fig8}. The significance of the Moon shadow is -139.6$\sigma$ in the energy band of 25-100 TeV and -86.8$\sigma$ above 100 TeV. The monthly results are presented in Fig \ref{fig8}. Since the Moon shadow is displaced along the R.A. direction by the geomagnetic field, the position in the Dec direction is usually adopted to check the array pointing accuracy. According to Fig \ref{fig8}, the pointing error in the Dec direction is much less than $0.1^\circ$ in all the months. A constant value is adopted to fit the points at 25-100 TeV, yielding that $\Delta Dec$= $-0.003^{\circ} \pm 0.004^{\circ}$. The corresponding value at $>$100 TeV is $\Delta Dec$= $0.001^{\circ} \pm 0.003^{\circ}$. The $\chi^2/ndf$ of the fitting is 36.09/23 and 26.55/23, respectively. This result is consistent with result achieved using the Crab Nebula. So $0.007^\circ$ and $0.004^\circ$ are the upper limits for combined systematic and statistical errors at 25-100 TeV and $>$100 TeV.

\begin{figure*}
\centering
\includegraphics[width=14cm,height=14cm]{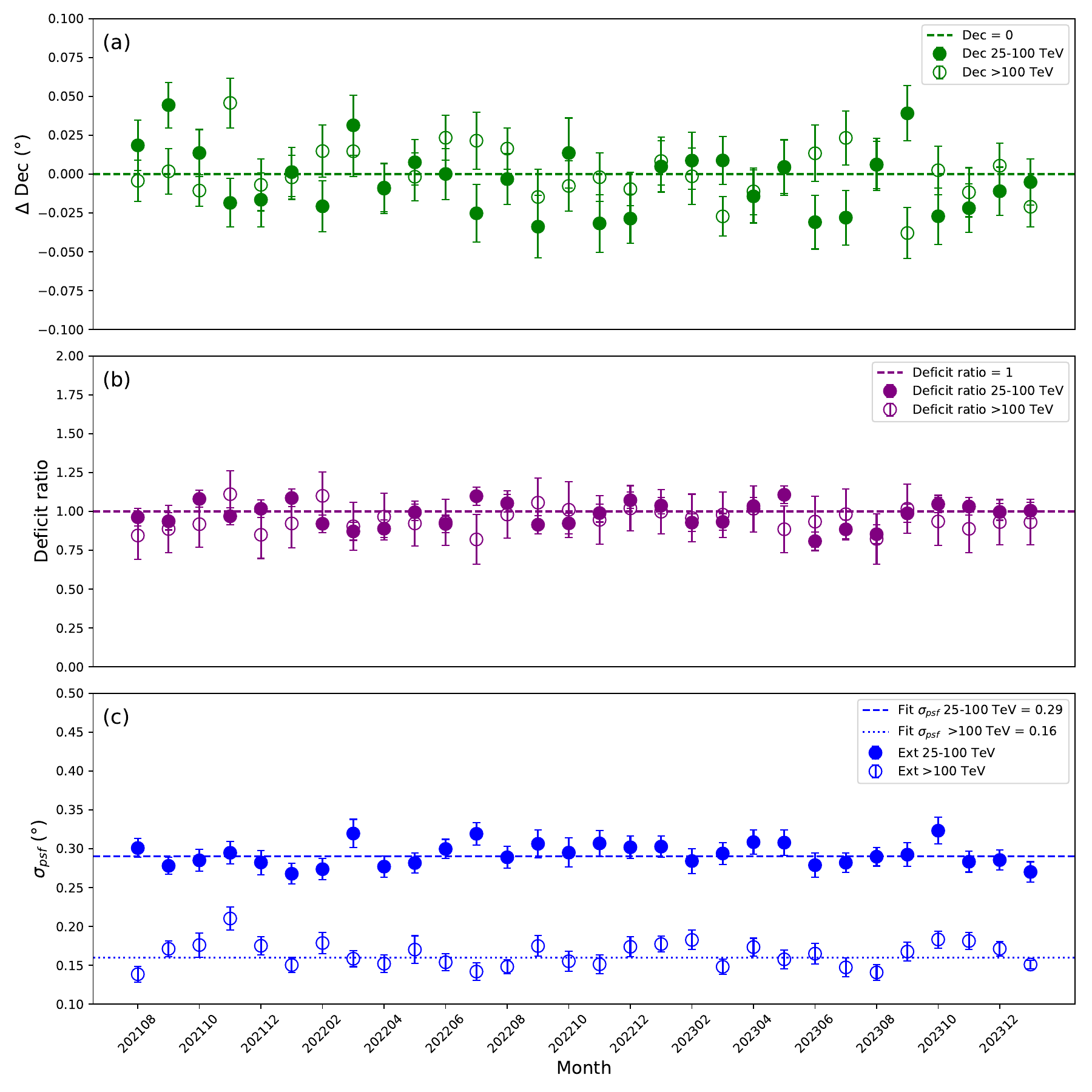}
\caption{Panel A: The monthly measured centroid of the Moon shadow relative to the expected Moon position in Dec direction as a function of time. The solid line shows a constant value that fits the centroid for all times. Panel B: The ratio of the measured Moon shadow deficit to the expected deficit as a function of time. Panel C: The monthly angular resolution obtained using the observation of the Moon shadow as a function of time.}
\label{fig9}
\end{figure*}

The cosmic ray Moon shadow is formed when the Moon blocks cosmic ray. Its expected deficit of events can be accurately estimated based on the distribution of background of events. Therefore, by comparing the measured deficit with the expected deficit, the data quality can be accurately evaluated. Fig \ref{fig9} also shows the ratio of the measured Moon shadow deficit to the expected deficit obtained each month. In this study, the expected deficit is the number of background events within $0.26^\circ$ of the shadow position, while the measured deficit of events was measured within $1.5^\circ$ of the shadow center for 25-100 TeV and $0.8^\circ$ for $>$ 100 TeV. This angle is larger than the angular resolution because of the impact of energy resolution. Cosmic ray with the same reconstructed energy band have a certain spread in their true energy, and cosmic ray with different energies are deflected differently by the Earth's magnetic field, resulting in an angular spread range larger than the angular resolution. According to Fig \ref{fig8}, we can see that the measured deficit is in good agreement with the expected value and remains stable over time. A constant value is adopted to fit the points at 25-100 TeV, yielding Deficit ratio = $0.98 \pm 0.02$ and the corresponding $\chi^2/ndf$ is 47.10/23. For the events at $>$100 TeV, the corresponding value is Deficit ratio = $0.99 \pm 0.04$ and the corresponding $\chi^2/ndf$ is 37.50/23. 

The intrinsic Moon shadow is an extended source like a disc with a radius of 0.26$^{\circ}$. Therefore, the effect of intrinsic extension needs to be removed from the angular distribution of the Moon shadow events when used to estimate the detector angular resolution. The monthly angular resolution results are also shown in Fig \ref{fig9}. A constant value is adopted to fit the points at 25-100 TeV, yielding that $\sigma_{PSF} $= $0.286^{\circ} \pm 0.003^{\circ}$ and the corresponding $\chi^2/ndf$ is 22.09/23. For the events at $>$100 TeV, the corresponding value is $\sigma_{PSF} $= $0.161^{\circ} \pm 0.003^{\circ}$ and $\chi^2/ndf$ is 41.75/23. Therefore, the angular resolution is stable month by month. So $0.289^\circ$ and $0.164^\circ$ are the upper limits for combined systematic and statistical errors at 25-100 TeV and $>$100 TeV.

\begin{figure*}
\centering
\includegraphics[width=10cm,height=8cm]{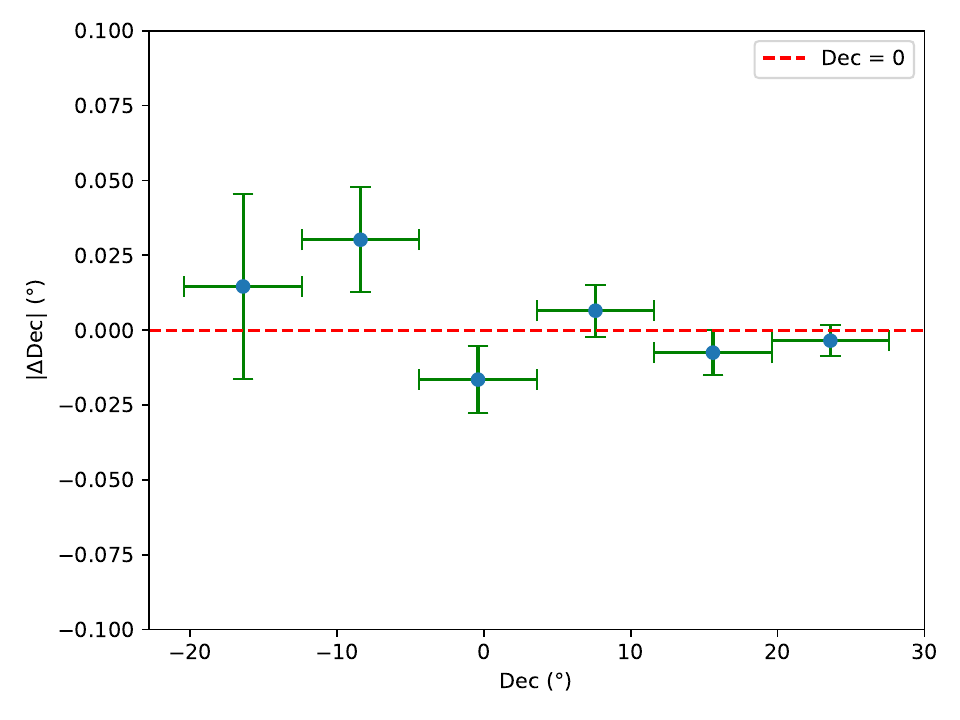}
\caption{The differences between the fitted and expected Dec of the Moon shadow vs. Dec. According to the Dec range of the Moon shadow in the KM2A field of view, Dec $-20.4^\circ$ to $27.6^\circ$ is selected, with each $8^\circ$ interval being treated as a bin. For LHAASO, the bins from left to right can be considered as corresponding to decreasing zenith angles.}
\label{fig10}
\end{figure*}


KM2A is a large field-of-view detection device that can observe the entire sky region in the Dec range from $-20^\circ$ to $80^\circ$ when events with a zenith angle less than $50^\circ$ are selected. If events with a zenith angle less than $60^\circ$ are chosen, it can observe the entire sky region in the Dec range from $-30^\circ$ to $90^\circ$. According to the first catalog of celestial sources detected by LHAASO, the gamma-ray sources detected by KM2A are widely distributed in the Dec direction. However, most of these sources are extended sources, and only the Crab Nebula with Dec = $22^\circ$ is suitable as a standard source to test detector performance. The pointing accuracy of KM2A for sources at different Dec has not been rigorously tested. Similar to KM2A, the HAWC array has found potential declination-dependent systematic biases in its pointing. The position of the Moon in the celestial coordinate system changes over time, and its Dec can vary monthly between $-28^\circ$ and $28^\circ$, providing us with the opportunity to test the pointing accuracy of the array in different Dec regions and very different zenith angles. Fig \ref{fig10} shows the measured position of the Moon shadow relative to the expected position in different Dec intervals. It can be seen that the measured position is consistent with the expected position, and no significant declination-dependent systematic bias has been found with an accuracy of $0.03^\circ$.

\section{Summary} 
As the most sensitive gamma-ray detector at energies above 10 TeV, KM2A has continuously monitored the overhead sky with the whole detector configuration and almost full duty cycle since July 20th, 2021. The KM2A comprises a large amount of detectors with more than 6000 units. All the detectors and their corresponding electronics are exposed to the harsh outdoor environment. To ensure the reliability of the data used for physical analysis, a three-level quality control system has been established for LHAASO-KM2A data. With this system, the data recorded by the abnormal detectors or in the abnormal data files can be filtered out. According to application to data collected during the period from August 2021 to July 2023, about 1.8\% of data files are filtered out. Within this system, the stability of the KM2A performance is also verified by monthly monitoring via the observations of the Crab Nebula and the Moon shadow. The pointing accuracy, angular resolution and detection efficiency have been very stable during the two years for operation. According to our result, $0.008^\circ$ and $0.023^\circ$ can be seen as conservative upper limits for combined systematic and statistical errors at 25-100 TeV and $>$100 TeV for the pointing accuracy on Crab Nebula observation, and $0.252^\circ$ and $0.151^\circ$ are the upper limits for combined systematic and statistical errors at 25-100 TeV and $>$100 TeV for the angular resolution on Crab Nebula observation. Similarly, $0.007^\circ$ and $0.004^\circ$ are the upper limits for combined systematic and statistical errors at 25-100 TeV and $>$100 TeV for the pointing accuracy on Moon shadow observation, and $0.289^\circ$ and $0.164^\circ$ are the upper limits for combined systematic and statistical errors at 25-100 TeV and $>$100 TeV for the angular resolution on Moon shadow observation.

In conclusion, we have established a quality control system for KM2A data, which effectively ensures data quality and monitors data stability. The LHAASO collaboration has conducted subsequent physical analyses based on the data generated by this system and has achieved numerous high-quality physics results. This quality control system could also applied to other EAS arrays.

\vspace{3mm}
\section*{Acknowledgments}
We would like to thank all staff members who work at the LHAASO site above 4400 meters above sea level year-round to maintain the detector and keep the water recycling system, electricity power supply and other components of the experiment operating smoothly. We are grateful to Chengdu Management Committee of Tianfu New Area for the constant financial support for research with LHAASO data. We deeply appreciate the computing and data service support provided by the National High Energy Physics Data Center for the data analysis in this paper. This research work is also supported by the following grants: the National Natural Science Foundation of China, NSFC No.12205314, No.12393851, No.12393854, No.12105301, No.12305120, No.12261160362, No.12105294, No.U1931201, No.12375107, No.12393852, No.12393853, the Natural Science Foundation of SiChuan Province of China, 2024NSFSC1372. In Thailand, support was provided from the NSRF via the Program Management Unit for Human Resources \& Institutional Development, Research and Innovation (B39G670013).

During the preparation of this work the authors used Chat GPT in order to fix syntax errors. After using this tool, the authors reviewed and edited the content as needed and takes full responsibility for the content of the publication.

\bibliographystyle{unsrt}
\bibliography{Ref}

\begin{thebibliography}{10}

\bibitem{greenwade93}
George~D. Greenwade.
\newblock The {C}omprehensive {T}ex {A}rchive {N}etwork ({CTAN}).
\newblock {\em TUGBoat}, 14(3):342--351, 1993.

\bibitem{2021Natur.594...33C}
Zhen {Cao}, F.~A. {Aharonian}, Q.~{An}, et~al.
\newblock {Ultrahigh-energy photons up to 1.4 petaelectronvolts from 12 {\ensuremath{\gamma}}-ray Galactic sources}.
\newblock {\em Nature}, 594(7861):33--36, June 2021.

\bibitem{2021ChPhC..45b5002A}
F.~{Aharonian}, Q.~{An}, {Axikegu}, et~al.
\newblock {Observation of the Crab Nebula with LHAASO-KM2A - a performance study}.
\newblock {\em Chinese Physics C}, 45(2):025002, February 2021.

\bibitem{2018APh...100...22L}
Hongkui {Lv}, Huihai {He}, Xiangdong {Sheng}, et~al.
\newblock {Calibration of the LHAASO-KM2A electromagnetic particle detectors using charged particles within the extensive air showers}.
\newblock {\em Astroparticle Physics}, 100:22--28, July 2018.

\bibitem{2014ChPhC..38b6001L}
Jia {Liu}, Xiang-Dong {Sheng}, Hui-Hai {He}, et~al.
\newblock {Performances and long-term stability of the LHAASO-KM2A prototype array}.
\newblock {\em Chinese Physics C}, 38(2):026001, February 2014.

\bibitem{2010ChPhC..34..249C}
Zhen {Cao}.
\newblock {A future project at tibet: the large high altitude air shower observatory (LHAASO)}.
\newblock {\em Chinese Physics C}, 34(2):249--252, February 2010.

\bibitem{2009NCimC..32e..19D}
A.~{De Angelis}.
\newblock {Very-high-energy gamma astrophysics}.
\newblock {\em Nuovo Cimento C Geophysics Space Physics C}, 32(5-6):19--26, September 2009.

\bibitem{2000RvMP...72..689N}
M.~{Nagano} and A.~A. {Watson}.
\newblock {Observations and implications of the ultrahigh-energy cosmic rays}.
\newblock {\em Reviews of Modern Physics}, 72(3):689--732, July 2000.

\bibitem{1947Natur.160..453L}
C.~M.~G. {Lattes}, G.~P.~S. {Occhialini}, and C.~F. {Powell}.
\newblock {Observations on the Tracks of Slow Mesons in Photographic Emulsions}.
\newblock {\em Nature}, 160(4066):453--456, October 1947.

\bibitem{1937PhRv...51..884N}
Seth~H. {Neddermeyer} and Carl~D. {Anderson}.
\newblock {Note on the Nature of Cosmic-Ray Particles}.
\newblock {\em Physical Review}, 51(10):884--886, May 1937.

\bibitem{1932PhRv...41..405A}
Carl~D. {Anderson}.
\newblock {Energies of Cosmic-Ray Particles}.
\newblock {\em Physical Review}, 41(4):405--421, August 1932.

\bibitem{Hess:1912srp}
Victor~F. Hess.
\newblock {\"Uber Beobachtungen der durchdringenden Strahlung bei sieben Freiballonfahrten}.
\newblock {\em Phys. Z.}, 13:1084--1091, 1912.

\bibitem{2023GRB}
{LHAASO Collaboration}.
\newblock Very high-energy gamma-ray emission beyond {10 TeV from GRB 221009A}.
\newblock {\em Science Advances}, 9(46), November 2023.

\bibitem{2023Cygnusss}
{LHAASO Collaboration}.
\newblock {An Ultrahigh-energy $\gamma$-ray Bubble Powered by a Super PeVatron}.
\newblock {\em arXiv e-prints}, page arXiv:2310.10100, October 2023.

\bibitem{2023Cygnus}
Zhen {Cao}, F.~{Aharonian}, Q.~{An}, {Axikegu}, et~al.
\newblock {An ultrahigh-energy {\ensuremath{\gamma}} -ray bubble powered by a super PeVatron}.
\newblock {\em Science Bulletin}, 69(4):449--457, February 2024.

\bibitem{2023diffusegamma}
Zhen Cao, F.~Aharonian, Q.~An, Axikegu, et~al.
\newblock Measurement of ultra-high-energy diffuse gamma-ray emission of the galactic plane from {10 TeV to 1 PeV with LHAASO-KM2A}.
\newblock {\em Physical Review Letters}, 131(15), October 2023.

\bibitem{2023firstcatalogss}
Zhen {Cao}, F.~{Aharonian}, Q.~{An}, {Axikegu}, et~al.
\newblock {The First LHAASO Catalog of Gamma-Ray Sources}.
\newblock {\em arXiv e-prints}, page arXiv:2305.17030, May 2023.

\bibitem{2023firstcatalog}
Zhen {Cao}, F.~{Aharonian}, Q.~{An}, {Axikegu}, et~al.
\newblock {The First LHAASO Catalog of Gamma-Ray Sources}.
\newblock {\em The Astrophysical Journal Supplement Series}, 271(1):25, March 2024.

\bibitem{2017ICRC}
Y.~{Nan} and S.~{Chen}.
\newblock {A study of the methods for signal significance estimation in ground-based gamma-ray detectors}.
\newblock In {\em 35th International Cosmic Ray Conference (ICRC2017)}, volume 301 of {\em International Cosmic Ray Conference}, page 879, July 2017.

\bibitem{2021Sci...373..425L}
{LHAASO Collaboration}, Zhen {Cao}, F.~{Aharonian}, et~al.
\newblock {Peta-electron volt gamma-ray emission from the Crab Nebula}.
\newblock {\em Science}, 373:425--430, July 2021.

\bibitem{2011Sci}
M.~{Tavani}, A.~{Bulgarelli}, V.~{Vittorini}, et~al.
\newblock {Discovery of Powerful Gamma-Ray Flares from the Crab Nebula}.
\newblock {\em Science}, 331(6018):736, February 2011.

\bibitem{2011Sci.}
A.~A. {Abdo}, M.~{Ackermann}, M.~{Ajello}, et~al.
\newblock {Gamma-Ray Flares from the Crab Nebula}.
\newblock {\em Science}, 331(6018):739, February 2011.

\bibitem{He2018}
Huihai He.
\newblock Design of the {LHAASO detectors}.
\newblock {\em Radiation Detection Technology and Methods}, 2, 06 2018.

\bibitem{2022ChPhC..46c0001M}
Xin-Hua {Ma}, Yu-Jiang {Bi}, Zhen {Cao}, et~al.
\newblock {Chapter 1 LHAASO Instruments and Detector technology}.
\newblock {\em Chinese Physics C}, 46(3):030001, March 2022.

\bibitem{WU201841}
Sha Wu, Liang Chen, Songzhan Chen, et~al.
\newblock Study of the trigger mode of {LHAASO-KM2A}.
\newblock {\em Astroparticle Physics}, 103:41--48, 2018.

\bibitem{Wang19}
Zheng Wang.
\newblock {\em LHAASO Electromagnetic particle detectors’ data quality monitoring Development of Intelligent monitoring system and Long term stability study}.
\newblock PhD thesis, Institute of High Energy Physics, Chinese Academy of Sciences, 2019.

\bibitem{Yan2023}
Shang Yanjun, Jin Weijun, Xiao Gang, Yang Peng, and He~Wantong.
\newblock Spatial variations of grain size of loose sediments in {Daocheng Haizi} mountain.
\newblock {\em Journal of Engineering Geology}, 31(5):1495--1506, 2023.

\bibitem{green2003}
D.~A. Green and F.~R. Stephenson.
\newblock {\em Supernovae and $\gamma$-ray Bursters}, volume 598.
\newblock Springer Verlag, Berlin, 2003.

\bibitem{YBJ2011}
B.~{Bartoli}, P.~{Bernardini}, X.~J. {Bi}, et~al.
\newblock {Observation of the cosmic ray moon shadowing effect with the ARGO-YBJ experiment}.
\newblock {\em Phys.Rev.D}, 84(2):022003, July 2011.

\bibitem{chsz2024}
Zhen Cao, Songzhan Chen, Ruoyu Liu, and Ruizhi Yang.
\newblock Ultra-high-energy gamma-ray astronomy.
\newblock {\em Annual Review of Nuclear and Particle Science}, 73(1):341--363, 2023.

\bibitem{PhysRevD.106.122004}
F.~Aharonian, Q.~An, Axikegu, L.~X. Bai, et~al.
\newblock Self-calibration of {LHAASO-KM2A} electromagnetic particle detectors using single particles within extensive air showers.
\newblock {\em Phys. Rev. D}, 106:122004, Dec 2022.

\bibitem{LV201822}
Hongkui Lv, Huihai He, Xiangdong Sheng, et~al.
\newblock Calibration of the {LHAASO-KM2A} electromagnetic particle detectors using charged particles within the extensive air showers.
\newblock {\em Astroparticle Physics}, 100:22--28, 2018.

\end{thebibliography}

\end{document}